\documentclass[a4paper,floatfix,twocolumn,prc,nofootinbib]{revtex4}
\usepackage{url}
\usepackage{graphicx}
\usepackage{epsfig}
\usepackage{gensymb}
\usepackage{amsmath}
\usepackage{amsfonts}
\usepackage{amssymb}
\usepackage{mathrsfs}
\newcommand{\vb}[1]{\boldsymbol{\mathrm {#1}}} 
\newcommand{\mat}[1]{\underline{\vb{#1}}} 
\newcommand{\ev}{\vb{e}}
\newcommand{\kv}{\vb{k}}
\newcommand{\rv}{\vb{r}}
\newcommand{\uv}{\vb{u}}
\newcommand{\vv}{\vb{v}}
\newcommand{\xiv}{\vb{\xi}}

\newcommand{\barphi}{\bar{\varphi}}
\newcommand{\nablav}{\vb{\nabla}}

\newcommand{\calH}{\mathscr{H}}
\newcommand{\calL}{\mathscr{L}}
\newcommand{\calT}{\mathscr{T}}
\newcommand{\calV}{\mathscr{V}}

\newcommand{\Eq}[1]{Eq.~(\ref{#1})}
\newcommand{\Eqs}[1]{Eqs.~(\ref{#1})}
\newcommand{\Fig}[1]{Fig.~\ref{#1}}
\newcommand{\Figs}[1]{Figs.~\ref{#1}}
\newcommand{\Ref}[1]{Ref.~\cite{#1}}
\newcommand{\Refs}[1]{Refs.~\cite{#1}}
\newcommand{\Sec}[1]{Sec.~\ref{#1}}
\newcommand{\Secs}[1]{Secs.~\ref{#1}}
\newcommand{\nucl}{np}
\newcommand{\RWS}{R_{\mathrm{WS}}}
\newcommand{\TF}{\mathrm{TF}}
\newcommand{\CS}{\mathrm{C+S}}
\newcommand{\E}[1]{\cdot 10^{#1}}
\usepackage{hyperref}

\begin{document}

\title{Long-wavelength phonons in the crystalline and pasta phases of
  neutron-star crusts}

\author{David Durel}
\email{david.durel@u-psud.fr}

\author{Michael Urban}
\email{urban@ipno.in2p3.fr}

\affiliation{Universit\'e Paris-Saclay, CNRS/IN2P3, IJCLab, 91405 Orsay, France}

\begin{abstract}
  We study the long-wavelength excitations of the inner crust of
  neutron stars, considering three phases: cubic crystal at low
  densities, rods and plates near the core-crust transition. To
  describe the phonons, we write an effective Lagrangian density in
  terms of the coarse-grained phase of the neutron superfluid gap and
  of the average displacement field of the clusters. The kinetic
  energy, including the entrainment of the neutron gas by the
  clusters, is obtained within a superfluid hydrodynamics
  approach. The potential energy is determined from a model where
  clusters and neutron gas are considered in phase coexistence,
  augmented by the elasticity of the lattice due to Coulomb and
  surface effects. All three phases show strong anisotropy, i.e.,
  angle dependence of the phonon velocities. Consequences for the
  specific heat at low temperature are discussed.
\end{abstract}

\maketitle

\section{Introduction}
In the inner crust of neutron stars, neutron-rich nuclei (clusters)
coexist with a gas of unbound neutrons and a degenerate electron gas
\cite{Chamel2008}. To minimize the Coulomb energy, the clusters form a
periodic lattice. Close to the transition to the neutron-star core,
the competition between Coulomb and surface energy leads to the
so-called ``pasta phases'': while the clusters are assumed to be
spherical at low density (crystalline phase), they merge with
increasing density to form rods (``spaghetti phase'') and then plates
(``lasagne phase'') \cite{Ravenhall1983}.

The neutrons in the inner crust are superfluid, which has important
effects for glitches and cooling of neutron stars. In particular, the
contribution of neutron quasiparticles to the specific heat of the
crust is strongly suppressed by pairing. Therefore, the dominant
contributions to the specific heat are those of the electrons, lattice
phonons, and superfluid phonons of the neutron gas
\cite{Page2012}. However, not all neutrons participate in the
superfluid motion of the neutron gas, because some are entrained by
the clusters. This entrainment effect, in addition to reducing the
superfluid density, leads also to a coupling between superfluid and
lattice phonons.

At low temperature, the long-wavelength phonons are most relevant for
the thermodynamic properties. In this article, we will only study
phonons of wavelengths greater than the lattice spacing. These phonons
can be described within an effective theory \cite{Cirigliano2011}
without having recourse to a microscopic model of the crust. However,
the parameters of this effective theory have to be determined from a
microscopic model. Here, we treat the relative motion between the gas
and the clusters within the hydrodynamic model of
Ref.~\cite{Martin2016}, which predicts a rather weak entrainment. We
consider the possibility that the superfluid and normal neutron
densities are not numbers but depend on the direction of the relative
velocity of neutrons and protons, as it is the case in the pasta
phases. This requires a generalization of the ``mixing'' term
\cite{Cirigliano2011}, coupling the superfluid phonons to the lattice
phonons.

We revisit also the elastic properties which determine the lattice
phonons. As in the case of condensed matter \cite{Fuchs1936}, the
anisotropy of the crystal leads to a splitting of the two transverse
phonons and to sound speeds that depend on the direction of the phonon
wave vector. The pasta phases are even more anisotropic. Like liquid
crystals in condensed matter, they can support shear stress only in
certain directions \cite{Pethick1998}. This results in a strong angle
dependence of the phonon velocities and changes qualitatively the
behavior of the specific heat at low temperature.

Our article is organized as follows. In \Sec{sec:coarse-grained}, we
review the basic idea of an effective theory as a result of
coarse-graining certain microscopic quantities. In
\Sec{sec:energycontributions}, we express the energy of the system in
terms of the coarse-grained variables, which will then lead us to the
effective Lagrangian in \Sec{sec:Lagrangian}. In \Sec{sec:results} we
present results for phonon energies and the specific heat. Finally, we
conclude in \Sec{sec:conclusions}.

Throughout the article, unless stated otherwise, we use units with
$\hbar = c = k_B = 1$, where $\hbar$ is the reduced Planck constant,
$c$ is the speed of light, and $k_B$ is the Boltzmann constant.
\section{Microscopic and coarse-grained quantities}
\label{sec:coarse-grained}
At a microscopic level, the neutron and proton densities $n_n$ and
$n_p$ vary at length scales much smaller than the periodicity of the
lattice. The same is true for the dynamical quantities. For example,
the motion of the neutrons is described by the phase of the superfluid
order parameter (gap) $\Delta = |\Delta|e^{2i\varphi}$, giving rise to
a velocity field $\vv_n = \nablav\varphi/m$, with $m$ the nucleon mass
(for convenience, we define $\varphi$ as one half of the phase). Both
$\vv$ and $\Delta$ can vary strongly inside one unit cell, even in the
case of a constant flow of the clusters through the gas
\cite{Martin2016}. In principle, also the proton velocity field
$\vv_p$ could vary on length scales much smaller than the periodicity
of the lattice, but this would correspond to a rather high-lying
internal excitation of the cluster which will be neglected here.

The basic idea of an effective theory is to describe long-wavelength
phenomena in terms of slowly varying quantities that can be obtained
by coarse-graining the microscopic quantities
\cite{Cirigliano2011}. Let us introduce the macroscopic neutron and
proton densities $\bar{n}_n$ and $\bar{n}_p$ which are obtained by
averaging the microscopic densities over a volume containing at least
one unit cell. For instance, in the simple phase-coexistence model
\cite{Martin2015} with constant density $n_{n,1}$ in the neutron gas
(volume $V_1$) and constant densities $n_{n,2}$ and $n_{p,2}$ inside
the clusters (volume $V_2$), one has $\bar{n}_n = (1-u)
n_{n,1}+un_{n,2}$ and $\bar{n}_p = u n_{p,2}$, where $u =
V_2/(V_1+V_2)$ is the volume fraction of the cluster (we assume that
there are no protons in the gas). Similarly, one can coarse-grain the
phase $\varphi$ to obtain a smoothly varying function $\barphi$. It
turns out that this averaged phase $\barphi$ determines the
macroscopic superfluid velocity, $\uv_n = \nablav\barphi/m$
\cite{Pethick2010,Martin2016}. Finally, as mentioned above, there is
not a big difference between the microscopic proton velocity $\vv_p$
and the average one $\uv_p$ as long as one does not consider
high-lying internal excitations of the clusters. In the case of pasta
phases, where the ``clusters'' are infinite in one (spaghetti) or two
(lasagne) directions, there exist of course also low-lying internal
excitations, which can be described by a slowly varying $\uv_p$.

The aim of the next subsections is to express the kinetic and
potential energies of the system entirely in terms of macroscopic
variables. Following \Ref{Cirigliano2011}, we will use as degrees of
freedom the coarse-grained phase $\barphi$ for the neutrons and the
average displacements $\xiv$ defined by $\dot{\xiv} = \uv_p$ for the
protons.

Note that in this work, we do not introduce any degrees of freedom
related to the electrons. That is, we assume that the electrons follow
instantly the motion of the protons such as to compensate the average
electric charge. This approximation requires that the wavelength of
the modes is large compared to the Thomas-Fermi screening length. In
doing so, we miss the damping of the modes which is to a large extent
generated by the electrons \cite{Kobyakov2017}.
\section{Contributions to the energy}
\label{sec:energycontributions}
\subsection{Kinetic energy density}
The kinetic energy density $\calT$ was determined in \cite{Martin2016}
using the superfluid hydrodynamics approach, where it was expressed as
a function of the velocities of the superfluid neutrons, $\uv_n$ and
of the protons, $\uv_p$ as follows:
\begin{equation}
\calT = \dfrac{1}{2} m \left( \uv_n\cdot
\mat{n}_n^s \uv_n + \uv_p\cdot
(\mat{n}_n^b + \bar{n}_p)\uv_p \right)\,.
\label{eq:kinetic-energy}
\end{equation}
The matrices $\mat{n}_n^s$ and $\mat{n}_n^b$ contain the densities of
superfluid and bound neutrons, respectively, along the different axes,
with $\mat{n}_n^s+\mat{n}_n^b=\bar{n}_n\mat{I}$, $\mat{I}$ being the
identity matrix.
\subsubsection{Crystalline phase}
In a crystal with cubic symmetry (such as the body-centered cubic
(BCC) crystal for the spherical clusters), these matrices reduce to
scalars and \Eq{eq:kinetic-energy} agrees with the expression given in
\Ref{Chamel2006} if one identifies $n_n^b$ with the neutron normal
density in the nomenclature of that reference. The computation of
$n_n^b$ and $n_n^s$ was presented in \cite{Martin2016} and it was
shown that to a very good approximation they can be obtained from the
analytical expressions for the effective mass of an isolated cluster
in an infinite neutron gas
\cite{Magierski2004a,Magierski2004b,Magierski2004c}. The corresponding
expression for $n_n^b$ reads
\begin{equation}
  n_n^b = u \, \frac{(1-\gamma)^{2}}{2\gamma + 1} \, n_{n,2}\,,
  \label{eq:nnb-crystal}
\end{equation}
with $\gamma = n_{n,1}/n_{n,2}$, and $n_n^s$ can be obtained from
$n_n^s = \bar{n}_n - n_n^b$.

\subsubsection{Spaghetti phase}
In the spaghetti phase (rods in $z$ direction), $\mat{n}_n^b$ and
$\mat{n}_n^s$ are diagonal in the $(x,y,z)$ coordinate system, but the
elements $n_{n,zz}^{b,s}$ are different from $n_{n,xx}^{b,s}$ and
$n_{n,yy}^{b,s}$. While in the numerical calculation of
\cite{Martin2016} a very weak anisotropy in the $xy$ plane was found,
this anisotropy must vanish exactly because of the discrete rotational
invariance of the hexagonal lattice under rotations by $60\degree$
around the $z$ axis, which was not recognized in \cite{Martin2016}.
Analogously to the crystalline case, an analytic formula has been
derived for the effective mass of an isolated rod in an infinite
neutron gas \cite{Martin2016}, and the corresponding expression for
the density of bound neutrons for a flow in the $xy$ plane reads
\begin{equation}
  n_{n,xx}^b = n_{n,yy}^b = u \, \frac{\left(1-\gamma\right)^2}{1+\gamma} \,
  n_{n,2}\,.
\end{equation}
For a flow in the direction of the rods, all neutrons are superfluid,
i.e.,
\begin{equation}
n_{n,zz}^b = 0.
\end{equation}

\subsubsection{Lasagne phase}
Let us finally consider the lasagne phase (plates parallel to the $xy$
plane). In this case, $\mat{n}_n^n$ and $\mat{n}_n^b$ are also
diagonal in the $(x,y,z)$ coordinate system with \cite{Martin2016}
\begin{gather}
n_{n,xx}^b = n_{n,yy}^b = 0\,,\\
n_{n,zz}^b = \frac{(1-\gamma)^2u(1-u)}{(\gamma u+1-u)}\,n_{n,2}\,. 
\end{gather}

\subsection{Potential energy density from the phase coexistence model}
\label{sec:phasecoexistence}
As discussed in \cite{Martin2015}, the microscopic equilibrium
densities satisfy to a good approximation the conditions of chemical
and mechanical equilibrium, i.e., $\mu_{a,1}=\mu_{a,2} \equiv \mu_a$
and $P_1 = P_2\equiv P$, where $\mu_{a,i}$ and $P_i$ are,
respectively, the chemical potential of species $a = n,p$ and the
pressure in phase $i=1$ (gas) or $2$ (cluster). The chemical
equilibrium condition for the protons can actually be omitted provided
that $\mu_p < 0$ (i.e., $n_{p,1}=0$). We assume that these equalities
remain valid under small variations of the density, whereas the
$\beta$ equilibrium $\mu_n=\mu_p+\mu_e$, which determines the volume
fraction $u$ in equilibrium, is not satisfied any more because the
weak processes $n\leftrightarrow p+e^-$ are too slow. The presence of
electrons as well as Coulomb and surface effects are neglected for the
moment.

Let us now use this simple phase coexistence model to compute the
variation of the energy in terms of variations of the macroscopic
variables (see also appendix of Ref.~\cite{Kobyakov2016}). Expanding
the chemical and mechanical equilibrium conditions to first order in
small variations around equilibrium, one finds ($a = n,p$)
\begin{gather}
  \delta n_{n,1} = \left.\frac{\partial n_n}{\partial\mu_n}\right|_1
    \delta\mu_n\,, \label{eq:deltann1}\\
    \delta n_{a,2} = \sum_{b=n,p} \gamma_b \left.
      \frac{\partial n_a}{\partial\mu_b} \right|_2 \delta\mu_n\,,
      \label{eq:deltana2}
\end{gather}
where we have introduced the abbreviations
\begin{equation}
  \gamma_n = 1\,,\qquad \gamma_p = \frac{n_{n,1}-n_{n,2}}{n_{p,2}}\,.
\end{equation}
The notation $|_i$ after the derivative means that the derivative is
evaluated at the equilibrium values in phase $i$. From
\Eqs{eq:deltann1} and \eqref{eq:deltana2} one sees that the neutron
and proton densities inside the clusters cannot oscillate
independently, but that they are tied to the oscillations of the
neutron density in the gas, uniquely determined by the neutron
chemical potential. Using the definitions of the average densities
$\bar{n}_a$, it is now straight-forward to obtain the relation
\begin{equation}
  \delta \mu_n = \frac{\delta\bar{n}_n+\gamma_p\delta\bar{n}_p}{\Gamma}
  \label{eq:deltamun}
\end{equation}
where
\begin{equation}
  \Gamma = (1-u)\left.\frac{\partial n_n}{\partial\mu_n}\right|_1
  + u \sum_{a,b=n,p}\gamma_a \gamma_b
  \left.\frac{\partial n_a}{\partial\mu_b}\right|_2\,.\label{eq:Gamma}
\end{equation}

Let us now consider
\begin{equation}
  \calV_{\nucl} = \bar{\varepsilon}_{\nucl}-\mu_n^{(0)}\bar{n}_n - 
    \mu_p^{(0)}\bar{n}_p\,,
  \label{eq:Vstrong}
\end{equation}
where $\bar{\varepsilon}_{\nucl} =
(1-u)\varepsilon_{\nucl}(n_{n,1},0)+u\varepsilon_{\nucl}(n_{n,2},n_{p,2})$
is the average energy density, determined from the nuclear energy
density functional $\varepsilon_{\nucl}(n_n,n_p)$ (in our case, the
Skyrme parameterization SLy4 \cite{Chabanat1998}), and $\mu_a^{(0)}$
are Lagrange parameters used to fix the equilibrium densities. The
equilibrium chemical potentials $\mu_a$ satisfy $\mu_a =
\mu_a^{(0)}$. Hence, the change in $\calV_{\nucl}$ is of second order
in the variations around equilibrium. Making use of
\Eqs{eq:deltann1}--(\ref{eq:Gamma}), one eventually obtains
\begin{equation}
  \delta\calV_{\nucl} = \frac{\Gamma}{2}\, (\delta\mu_n)^2\,.
\end{equation}
This expression is a special case of Eq.~(12) of \cite{Kobyakov2016}
which states that the total energy change $\delta\calV$ can be written
as a term $\propto \delta\mu_n^2$ and a term
$\propto(\delta\bar{n}_p)^2$ with no cross term $\propto
\delta\mu_n\delta \bar{n}_p$. In the present case, it turns out that
the term $\propto (\delta\bar{n}_p)^2$ is absent: If one changes the
density of clusters (i.e., $u$), without changing the microscopic
densities $n_{a,i}$ inside the gas or the clusters, the total energy
does not change. This unphysical property of the model will be
corrected when we include the electron contribution.
\subsection{Electron contribution}
\label{sec:electrons}
To include the electrons, we replace \Eq{eq:Vstrong} by
\begin{equation}
\calV_{\nucl}+\calV_e = \bar{\varepsilon}_{\nucl}+\bar{\varepsilon}_e-\mu_n^{(0)}
  (\bar{n}_n + \bar{n}_p)\,.
\end{equation}
In writing this equation, we made use of the relation $\mu_n^{(0)} =
\mu_p^{(0)}+\mu_e^{(0)}$, which is a consequence of $\beta$
equilibrium in the ground state. Furthermore, as already mentioned in
the end of \Sec{sec:coarse-grained}, we assumed that the electrons
follow instantly the protons to maintain the average charge
neutrality: $\bar{n}_e = \bar{n}_p$. In contrast to the microscopic
neutron and proton densities $n_n$ and $n_p$, the microscopic electron
density $n_e$ is to a very good approximation constant over a unit
cell, since the screening length is larger than the periodicity of the
lattice. Hence, we have $n_e = \bar{n}_e = \bar{n}_p$ and the electron
energy density is given by $\bar{\varepsilon}_e =
\varepsilon_e(\bar{n}_p)$.

For small oscillations around equilibrium, the linear term vanishes
again and the quadratic term is given by
\begin{equation}
  \delta\calV_e
  = \frac{K}{2} \left(\frac{\delta\bar{n}_p}{\bar{n}_p}\right)^2
  = \frac{K}{2} (\nablav\cdot\xiv)^2\,,
\end{equation}
with the bulk modulus $K = \bar{n}_p^2 (\partial \mu_e / \partial
n_e)_{n_e = \bar{n}_p}$. If one neglects the electron mass, the
electron chemical potential $\mu_e = \partial \varepsilon_e/\partial
n_e$ reads $\mu_e = \hbar c(3\pi^2n_e)^{1/3}$ and therefore
\begin{equation}
  K = \Big(\frac{\pi}{3}\Big)^{2/3} \hbar c\, \bar{n}_p^{4/3}\,.
\end{equation}

Note that our $K$ is different from the bulk modulus $\tilde{K}$
defined in \Ref{Kobyakov2013} since we vary $\bar{n}_p$ keeping
$\mu_n$ constant, while in \Ref{Kobyakov2013} $\bar{n}_p$ is varied
keeping $\bar{n}_n$ constant.
\subsection{Coulomb and surface energy: elastic constants}
\label{sec:elasticity}
So far, the energy depends only on the dilatation or compression of
the lattice, determined by $\nablav\cdot\xiv$. To describe the
elasticity of the crust, i.e., the energy cost of shear deformations,
it is necessary to include also the Coulomb and surface energy
$\varepsilon_{\CS}$. Note that the microscopic equilibrium quantities
$n_{n,1}$, $n_{n,2}$, $n_{p,2}$, $u$, etc., are in principle
determined by minimizing the total energy including
$\varepsilon_{\CS}$, i.e., \cite{Martin2015},
\begin{equation}
  \calV = \bar{\varepsilon}_{\nucl}+\varepsilon_e+\varepsilon_{\CS}
  -\mu_n^{(0)}(\bar{n}_n+\bar{n}_p)\,.
\end{equation}
The presence of $\varepsilon_{\CS}$ leads to small corrections to the
phase coexistence conditions of \Sec{sec:phasecoexistence}, but with
the corrected equilibrium values, $\calV$ is stationary again. Hence,
the variation of $\calV$ in the case of small oscillations around
equilibrium is determined by the second derivatives, as in the case
without Coulomb and surface energies discussed in
\Secs{sec:phasecoexistence} and \ref{sec:electrons}.
\subsubsection{Lasagne phase}
Let us start with the lasagne phase. In this case, we have
\begin{equation}
  \varepsilon_{\CS} = \frac{2\pi}{3} e^2 n_{p,2}^2\RWS^2 u^2(1-u)^2
    +\frac{\sigma}{\RWS}\,,
\end{equation}
where $\RWS=L/2$ is one half of the periodicity of the structure, and
$\sigma$ is the surface tension. The surface tension may be eliminated
using the equilibrium condition $\partial\varepsilon_{\CS}/\partial
\RWS=0$, which gives $\sigma = (4\pi/3) e^2 n_{p,2}^2
\RWS^3u^2(1-u)^2$ (with $n_{p,2}$, $\RWS$, and $u$ being the
equilibrium values).

\begin{figure}
\includegraphics[width=7cm]{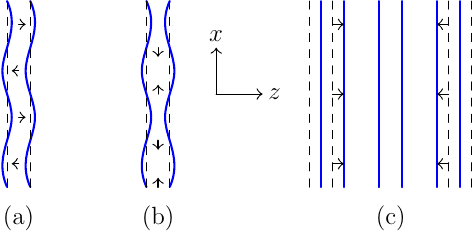}
\caption{\label{fig:deformation-lasagne}Illustration of different
  types of deformations leading to a change of Coulomb and surface
  energies in the lasagne phase. The small arrows indicate the
  displacement field $\xiv$. (a) spatially varying shear in the $xz$
  plane (cf. \Eq{eq:k4-lasagne}), (b) compression by displacing
  protons inside the lasagne plates ($B_{11}$ term in
  \Eq{eq:pasta-elasticity}), (c) compression by changing the distance
  between the plates ($B_{33}$ term in \Eq{eq:pasta-elasticity}).}
\end{figure}

Neglecting possible coupling terms involving $\delta\mu_n$ and $\xiv$,
we consider $\mu_n$ and therefore also $n_{p,2}$ constant, and vary
only $u$ and $\RWS$.\footnote{In \Ref{Kobyakov2016} it is shown that
  there is no cross term $\propto\delta\mu_n\nablav\cdot\xiv$. But
  through the $n_{p,2}$ dependence of the Coulomb energy and the
  density dependence of $\sigma$ there might be an anisotropic
  coupling between some shear deformations and $\delta\mu_n$, which we
  neglect.} Since all first-order derivatives vanish, we may simply
add to $\delta\calV_{\nucl}+\delta\calV_e$ the second-order variation
of $\varepsilon_{\CS}$,
\begin{multline}
  \delta\calV_{\CS,1} =
  \frac{\partial^2\varepsilon_{\CS}}{\partial \RWS^2}
    \frac{(\delta \RWS)^2}{2}
  +\frac{\partial^2\varepsilon_{\CS}}{\partial \RWS\partial u}
    \delta \RWS \delta u\\
  +\frac{\partial^2\varepsilon_{\CS}}{\partial u^2}
    \frac{(\delta u)^2}{2}\,.
  \label{eq:secondorderuRWS}
\end{multline}
In the lasagne phase, one has $\delta \RWS = \RWS \partial_z \xi_z$
and $\delta u = -u \nablav\cdot\xiv$. The different roles of the
displacement fields in the $xy$ plane and in the direction
prependicular to it are illustrated in
\Fig{fig:deformation-lasagne}(b)
and (c). The energy can now be written in the form
\begin{multline}
  \delta\calV_{\CS,1} = \frac{B_{11}}{2}(\partial_x\xi_x+\partial_y\xi_y)^2
  + B_{13} (\partial_x\xi_x+\partial_y\xi_y) (\partial_z \xi_z)\\
  + \frac{B_{33}}{2}(\partial_z \xi_z)^2\,,
    \label{eq:pasta-elasticity}
\end{multline}
with
\begin{gather}
  B_{11} = \tfrac{4}{3} (1-6u+6u^2) C_0\,,\\
  B_{13} = \tfrac{4}{3} (-1+2u^2) C_0\,,\\
  B_{33} = \tfrac{4}{3} u^2 C_0\,,
\end{gather}
where we have introduced the abbreviation
\begin{equation}
  C_0 = \pi e^2 \bar{n}_p^2 \RWS^2\,.
\end{equation}
Our result differs from the one given in \Ref{Pethick1998} where only
the first term of \Eq{eq:secondorderuRWS} was taken into account. Note
that the combination $K_{\CS} = (4B_{11}+4B_{13}+B_{33})/9$ represents
a (negative) correction to the bulk modulus, but it is much smaller
than the contribution of the electrons, $K$.

From the particular geometry of the lasagne phase it is clear that
shear deformations in the $xy$ plane do not change the energy. As
noticed in \Ref{Pethick1998}, spatially constant shear deformations in
the $xz$ (or $yz$) plane contribute only at fourth order in their
amplitude (terms $\propto (\partial_x \xi_z)^4$ etc.). However,
spatially varying shear deformations in the $xz$ (or $yz$) plane (cf. 
\Fig{fig:deformation-lasagne}(a)) contribute already at second order
in the amplitude. The corresponding energy can be written as
\cite{Pethick1998}
\begin{equation}
 \delta\calV_{\CS,2} =
 \frac{K_1}{2}(\partial_x^2\xi_z+\partial_y^2\xi_z)^2\,,
\label{eq:k4-lasagne}
\end{equation}
To find the coefficient $K_1$, one considers, e.g., a displacement
field $\xi_z = \xi_0 \cos kx$, and calculates the change in energy to
second order in $\xi_0$. The $k^2\xi_0^2$ terms coming from the
Coulomb and the surface energies cancel, and the leading non-vanishing
term is proportional to $k^4\xi_0^2$. The coefficient of this term can
be identified with $K_1/4$, and one obtains \cite{Pethick1998}
\begin{equation}
  K_1 = \tfrac{4}{45}\pi e^2\bar{n}_p^2 \RWS^4 (1-u)^2(1+2u-2u^2)\,.
\end{equation}

\subsubsection{Spaghetti phase}
Now let us discuss the spaghetti phase. The exact calculation of the
Coulomb energy is quite involved in this case, but, except for the
energy due to shear deformations in the $xy$ plane, we may use the
Wigner-Seitz approximation as a first estimate: the unit cell in the
$xy$ plane, which is a rhombus with side length $L$ and angle
$60\degree$, is replaced by a circle of radius $\RWS$ with the same
area (i.e., $\RWS = 3^{1/4}L/\sqrt{2\pi}$). In this approximation,
the Coulomb energy is readily calculated. The sum of Coulomb and
surface energies per volume is
\begin{equation}
  \varepsilon_{\CS} = \frac{\pi}{2} e^2 n_{p,2}^2\RWS^2 u^{2} (-1+u-\ln u)
  +\frac{2\sqrt{u}\sigma}{\RWS}\,.
  \label{eq:ecsspaghetti}
\end{equation}
Again, the condition $\partial \varepsilon_{\CS}/\partial \RWS=0$
allows one to express $\sigma$ in terms of the equilibrium values of
$n_{p,2}$, $u$, and $\RWS$. The calculation of the energy variation
under dilatations in the plane perpendicular to the rods ($xy$) or in
the direction of the rods ($z$) is completely analogous to the case of
the lasagne phase (in \Fig{fig:deformation-lasagne} one only has to exchange
 $x$ and $z$ directions). In the spaghetti phase, one has $\delta u = -u
\nablav\cdot\xiv$ and $\delta \RWS =
\frac{1}{2} \RWS \nablav_\perp \xiv_\perp$ (with $\xiv_\perp$ and $ \nablav_\perp $ 
the projections of $\xiv$ and $ \nablav $ onto the $xy$ plane). Inserting 
\Eq{eq:ecsspaghetti} into
\Eq{eq:secondorderuRWS}, the energy change is again of the form of
\Eq{eq:pasta-elasticity}, only the coefficients are different:
\begin{gather}
  B_{11} = \tfrac{1}{4}(-2+4u) C_0\,,\\
  B_{13} = \tfrac{1}{4}(-4+6u) C_0\,,\\
  B_{33} = \tfrac{1}{4}(-9+11u-3\ln u) C_0 \, .
\end{gather}

Analogous to the case of the lasagne phase, a constant shear
deformation in the $xz$ (or $yz$) plane does not change the energy at
second order in the amplitude, but a shear deformation that oscillates
as a function of $z$ does. The corresponding energy is now written in
the form \cite{Pethick1998}
\begin{equation}
 \delta\calV_{\CS,2} =
 \frac{K_3}{2}(\partial_z^2\xiv_\perp)^2\,.
\label{eq:k4-spaghetti}
\end{equation}
To find the coefficient $K_3$, we follow again \Ref{Pethick1998} and
calculate the Coulomb energy for cylindrical spaghetti arranged in a
hexagonal lattice, which are displaced by $\xiv = \xiv_0 \cos kz$
($\xiv_0$ lying in the $xy$ plane):\footnote{Eq.~(12) in
  \Ref{Pethick1998} is missing a factor of $u = R^2/\RWS^2$
  (denoted $w = r_N^2/r_c^2$ there).}
\begin{equation}
  \varepsilon_C = \frac{8\pi e^2 n_{p,2}^2 u}{\RWS^2}
  \sideset{}{'}\sum_{lmn}
  \frac{[J_n(\kv_{lm}\cdot\xiv_0) J_1(k_{lm}R)]^2}{k_{lm}^2(k_{lm}^2+n^2k^2)}\,.
  \label{eq:eClasagnelattice}
\end{equation}
The prime indicates that the term $l=m=n=0$ is excluded. The
$\kv_{lm} = l \vb{b}_1+m\vb{b}_2$ are reciprocal lattice vectors,
$\vb{b}_1$ and $\vb{b}_2$ have length $b=3^{-1/4}\sqrt{8\pi}/\RWS$
and the angle between them is $60\degree$, such that $k_{lm}^2 =
b^2(l^2+m^2+l m)$. To find $K_3$, we expand \Eq{eq:eClasagnelattice} in
$k$ and $\xiv_0$ and retain only the term $\propto k^4\xi_0^2$, since
the term $\propto k^2\xi_0^2$ must be cancelled by the surface
energy. Only the terms $n=\pm 1$ contribute and we obtain
\begin{multline}
  K_3 \approx \frac{e^2 n_{p,2}^2 \RWS^4 u}{32\sqrt{3}\pi^2}
  \left(27 [J_1(3^{-1/4}\sqrt{8\pi u})]^2\right.\\
  +\left.[J_1(3^{1/4}\sqrt{8\pi u})]^2+\cdots\right)\,,
  \label{eq:K3analytic}
\end{multline}
where we have kept only the largest and second largest terms in the
sum over $l$ and $m$ since the sum converges extremely well.

Let us now turn to the shear modulus for shear in the $xy$ plane,
which was of course absent in the lasagne phase while in the spaghetti
phase it exists. The corresponding energy can be written as
\begin{equation}
  \delta\calV_{\CS,3} = \frac{C_{66}}{2}\left[(\partial_x\xi_x-\partial_y\xi_y)^2
  +(\partial_y\xi_x+\partial_x\xi_y)^2\right]\,.
\label{eq:spaghetti-shearxy}
\end{equation}
The invariance of the hexagonal lattice under rotations by
$60\degree$ implies that the same elastic constant $C_{66}$ appears in
front of the two terms. To obtain $C_{66}$, we again have to start
from the exact expression of the Coulomb energy, but now for a finite
shear $\partial_y\xi_x\equiv \xi_{xy}$ which changes the reciprocal
lattice vectors into $\vb{b}_1' = \vb{b}_1$ and
$\vb{b}_2'=\vb{b}_2+(\sqrt{3}\xi_{xy}/2)\vb{b}_1$ (we choose the
coordinate system such that $\vb{b}_1$ points in $y$ direction). Then
the Coulomb energy becomes
\begin{equation}
  \varepsilon_C= \frac{8\pi e^2 n_{p,2}^2 u}{\RWS^2}
  \sideset{}{'}\sum_{lm} \frac{[J_1(k'_{lm} R)]^2}{k^{\prime\,4}_{lm}}
\end{equation}
with $\kv'_{lm} = l\vb{b}'_1+m\vb{b}'_2$. Expanding this expression
up to order $\xi_{xy}^2$, keeping only the quadratic term, and using
the invariance of the lattice under rotations by multiples of
$60\degree$, one eventually obtains (see Appendix \ref{app:sumlm})
\begin{multline}
  C_{66} = 2\pi e^2 n_{p,2}^2 u^2
  \sum_{lm} \Big(\frac{[J_2(k_{lm} R)]^2}{k^2_{lm}}\\
  +\frac{4 J_1(k_{lm} R)J_2(k_{lm} R)}{k_{lm}^3R}
  -\frac{[J_1(k_{lm} R)]^2}{k^2_{lm}}\Big)\,.
  \label{eq:C66lattice}
\end{multline}
As explained in Appendix \ref{app:sumlm}, one can rewrite this sum as
an integral in coordinate space which can be done analytically, with
the result
\begin{equation}
  C_{66} = \tfrac{1}{4} C_0\,.
  \label{eq:C66analytic}
\end{equation}

So far, the elastic coefficients have been calculated without
screening of the Coulomb interaction by the electrons. In order to
include this effect, one has to replace $k_{lm}^2+n^2k^2$ in the
denominator of \Eq{eq:eClasagnelattice} by
$k_{lm}^2+n^2k^2+k_{\TF}^2$, with $k_{\TF}^2 =
(4e^2/\pi)(3\pi^2n_e)^{2/3}$. Repeating
the steps described before, performing the remaining summations over
$l$ and $m$ numerically, we found that the screening effect is weak
and may be neglected. We note also that our analytical results
(\ref{eq:K3analytic}) and (\ref{eq:C66analytic}) agree with the
numerical results shown in Fig.~2 of \Ref{Pethick1998}.

\subsubsection{Crystalline phase}
Finally, let us consider the crystalline phase. For a cubic crystal,
there are only three independent elastic constants, say, $C_{11}$,
$C_{12}$, and $C_{44}$ \cite{AshcroftMermin}. The combination
$(C_{11}+2C_{12})/3$ is the bulk modulus and we assume that it is
dominated by the electron contribution $K$ discussed in
\Sec{sec:electrons} so that the Coulomb contribution to it can be
neglected. The energy due to shear deformations is written as
\begin{multline}
  \delta\calV_{\CS} = \frac{C_{11}-C_{12}}{2}\sum_{i=x,y,z}
  \left(\partial_i\xi_i-\frac{\nablav\cdot\xiv}{3}\right)^2\\
  +C_{44}\sum_{i\neq j}
  \left(\frac{\partial_i\xi_j+\partial_j\xi_i}{2}\right)^2\,.
  \label{eq:elastic_bcc}
\end{multline}
The constants $C_{11}-C_{12}$ and $C_{44}$ were calculated, e.g., in
\Refs{Fuchs1936,Ogata1990}:\footnote{In \Ref{Ogata1990}, the numbers
  are given in units of $n_I (Ze)^2/\RWS = 4C_0/3$, while in
  \Ref{Fuchs1936}, they are given in units of $n_I (Ze)^2/(2L) =
  (3/\pi)^{1/3}C_0/3$ and $(3/2\pi)^{1/3}C_0/3$, respectively, for the
  BCC and face-centered cubic (FCC) lattice, with $n_I$ the number
  density of ions (clusters), $Ze$ the cluster charge, and $L$ the
  lattice constant. Note that the BCC and FCC unit cells of volume
  $L^3$ contain two and four clusters, respectively.}
\begin{equation}
  C_{11}-C_{12} = 0.06545\, C_0\,,\quad C_{44} = 0.2437\, C_0\,.
  \label{eq:elastic_bcc_coefficients}
\end{equation}
Screening corrections to these numbers were computed in
\Ref{Baiko2015}, but in the inner crust they are weak.

From \Eq{eq:elastic_bcc_coefficients} one sees that the anisotropy of
the BCC crystal is very strong (an isotropic material has
$C_{11}-C_{12} = 2C_{44}$). Since the crust is probably a polycrystal
made of many crystallites having random orientations, one often uses
an effective shear modulus obtained by a suitable averaging
\cite{Kobyakov2015}. This procedure seems very reasonable for the
description of macroscopic phenomena such as vibrations of the star,
but it is probably not adequate for the description of phonons whose
wavelength we assume to be large compared to the lattice constant $L$
but not necessarily large compared to the size of the crystallites. We
therefore keep the anisotropic form of \Eq{eq:elastic_bcc}, as we did
for the pasta phases, where the same argument applies.

\section{Lagrangian density}
\label{sec:Lagrangian}
\subsection{Lagrangian density for pure neutron matter}
\label{sec:LPNM}
Before writing the Lagrangian density describing the inner crust of
neutron stars, let us consider as a pedagogical example the simpler
case of pure neutron matter. In terms of the phase $\varphi$ as degree
of freedom, the Lagrangian of a one-component superfluid reads
\cite{Son2006} (see also \cite{Greiter1989})
\begin{equation}
  \calL = P \left(\mu_n^{(0)} - \dot{\varphi} -\frac{(\nablav\varphi)^2}{2m}
    \right)\,,
\label{eq:SonWingate}
\end{equation}
where $P(\mu_n)$ is the pressure of neutron matter und $\mu_n$ the
chemical potential. The superscript $(0)$ indicates equilibrium
quantities. 

The conjugate momentum to the field $\varphi$ is given by
$\partial\calL/\partial\dot{\varphi} = -\partial P/\partial\mu_n =
-n_n$, so that the Hamiltonian takes the expected form
\begin{equation}
  \calH = -n_n\dot{\varphi}-\calL =
  \varepsilon-\mu^{(0)}n_n+\frac{n_n^{(0)}}{2m}(\nablav\varphi)^2+\cdots\,,
\end{equation}
where the dots stand for terms of third or higher order in $\varphi$
that will be neglected, and we have used $P = \mu_n n_n-\varepsilon$.

In the notation of \Sec{sec:energycontributions}, we could write
$\calH = \calV+\calT$. The Lagrangian, however, would not be given by
$\calT-\calV$, but by $\calL = P^{(0)}-n_n^{(0)}
\dot{\varphi}+\delta\calV-\calT$. There is a linear term, which does
not contribute to the equations of motion, and the roles of $\calT$
and $\calV$ are exchanged, because the kinetic energy $\calT$ contains
only spatial derivatives of $\varphi$ while the time derivatives of
$\varphi$ enter the potential energy $\calV$.

\subsection{Lagrangian density for the inner crust}
Now we want to write down the Lagrangian for the inner crust in terms
of the fields $\barphi$ and $\xiv$. As we have seen in the example of
pure neutron matter, the Lagrangian is not given by
$\calT-\calV$. Therefore we start from the Hamiltonian which we write
as $\calH=\calH^{(0)}+\calT+\delta\calV$, with $\calT$ and
$\delta\calV$ from \Sec{sec:energycontributions}, replacing $\uv_n =
\nablav\barphi/m$, $\uv_p = \dot{\xiv}$, and $\delta\mu_n =
-\dot{\barphi}$. As before, we keep only terms up to second order in
the deviations from equilibrium. A Lagrangian that corresponds to this
Hamiltonian is given by
\begin{multline}
  \calL = -\calH^{(0)}
  +\frac{\Gamma}{2}\dot{\barphi}^2
  -\frac{1}{2m}(\nablav\barphi)\cdot\mat{n}_n^s\nablav\barphi
  -\dot{\barphi}\nablav\cdot\mat{\alpha}\xiv\\
  +\frac{m}{2}\dot{\xiv}\cdot(\mat{n}_n^b+\bar{n}_p\mat{I})\dot{\xiv}
  -\delta\calV_e-\delta\calV_{\CS}\,.
  \label{eq:Lcrust}
\end{multline}
To this Lagrangian, one could add a term $\propto\dot{\barphi}$ as it
was present in pure neutron matter, but such a term does not affect
the equations of motion. Furthermore, one could add a term
$\dot{\xiv}\cdot\mat{\beta}\nablav\barphi$, but up to total
derivatives such a term can be absorbed in the term
$-\dot{\barphi}\nablav\cdot\mat{\alpha}\xiv$ already present in
\Eq{eq:Lcrust}.

In order to determine $\mat{\alpha}$, one has to use an additional
information, namely the conservation of the neutron number, which can
be expressed as a continuity equation
\begin{equation}
\delta\dot{\bar{n}}_n + \nablav\cdot\left(\mat{n}_n^s\frac{\nablav\barphi}{m}
  +\mat{n}_n^b\dot{\xiv}\right) = 0\,.\label{eq:continuity}
\end{equation}
From \Eq{eq:deltamun}, one has $\delta\bar{n}_n =
-\Gamma\dot{\barphi}+\gamma_p\bar{n}_p\nablav\cdot\xiv$. Inserting now
the equation of motion
\begin{equation}
  \left(\frac{\partial\calL}{\partial\dot{\barphi}}\right)^{\!\!\!\mbox{$\cdot$}}
  +\nablav\cdot\frac{\partial\calL}{\partial(\nablav\barphi)}
  = \Gamma \ddot{\barphi}-\nablav\cdot\left(\mat{\alpha}\dot{\xiv}
  +\mat{n}_n^s \frac{\nablav\barphi}{m}\right) = 0\,,
  \label{eq:EulerLagrangephi}
\end{equation}
one finds
\begin{equation}
  \mat{\alpha} = \gamma_p \bar{n}_p\mat{I}+\mat{n}_n^b\,.
\end{equation}

The term $-\dot{\barphi}\nablav\cdot\mat{\alpha}\xiv$ generalizes the
mixing term introduced by Cirigliano \textit{et al.}
\cite{Cirigliano2011} for the isotropic case (cubic crystal), in which
it reduces to $-\alpha\dot{\barphi}\nablav\cdot\xiv$. Our $\alpha$
corresponds to $f_\phi\sqrt{\rho}g_{\text{mix}}$ in the notation of
\Ref{Cirigliano2011}. There, its value was determined by imposing
gauge invariance, which is equivalent to imposing the validity of the
continuity equation. In Eq.~(62) of \Ref{Cirigliano2011}, our factor
$\gamma_p$, which is defined in the simplified model of constant
densities in both phases with a sharp interface, is replaced by the
more general expression $(\partial\bar{n}_n/\partial\bar{n}_p)_{\mu_n}$.

\section{Results}
\label{sec:results}
In order to obtain the phonon dispersion relations, we write down the
Euler-Lagrange equations, i.e., \Eq{eq:EulerLagrangephi} and the
analogous equations of motion for the components of $\xiv$,
\begin{equation}
  \left(\frac{\partial\calL}{\partial\dot{\xi_i}}\right)^{\!\!\!\mbox{$\cdot$}}
  +\sum_j\partial_j\frac{\partial\calL}{\partial(\partial_j\xi_i)}
  -\sum_{jk}\partial_j\partial_k
    \frac{\partial\calL}{\partial(\partial_j\partial_k\xi_i)}=0\,.
  \label{eq:EulerLagrangexi}
\end{equation}
Assuming $\barphi = \barphi_0 e^{i (\kv \cdot \rv - \omega t)}$ and
$\xiv = \xiv_0 e^{i (\kv \cdot \rv - \omega t)}$, one can reduce the
problem to an algebraic matrix equation (see Appendix \ref{app:matrices}). 
For a given wave vector $\kv$, it is straight-forward to find the eigenvalues 
$\omega_i(\kv)$ and the corresponding eigenvectors $(\barphi_{0i},\xiv_{0i})$.

\subsection{Lasagne phase}
Because of the invariance of the lasagne phase with respect to
rotations around the $z$ axis, one may consider without loss of
generality the case $k_x = k\sin\theta$, $k_y = 0$, and $k_z =
k\cos\theta$. Furthermore, since the lasagne phase does not support
any shear stress in the $xy$ plane, a transverse mode having $\xiv$ in
$y$ direction has $\omega=0$. It is therefore sufficient to consider
only the degrees of freedom $\barphi$, $\xi_x$ and $\xi_z$, and there
are only three eigenmodes with $\omega>0$.

\begin{figure}
\includegraphics[width=7cm]{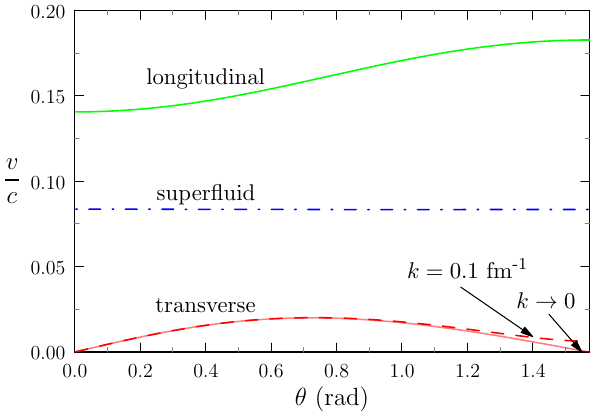}
\caption{\label{fig:sound-lasagne}Angle dependence of the velocities
  of the three phonons in the lasagne phase (parameters see Table
  \ref{tab:parameters}).}
\end{figure}

In \Fig{fig:sound-lasagne},
\begin{table}
  \caption{\label{tab:parameters} Parameters used in
    \Figs{fig:sound-lasagne}--\ref{fig:sound-crystal} (from the
    Extended Thomas-Fermi (ETF) calculation of \Ref{Martin2015}).}
  \begin{ruledtabular}
    \begin{tabular}{ccccc}
               & & lasagne & spaghetti & BCC crystal\\ \hline
      $\rho_B$ & (g/cm$^3$) & $1.27\E{14}$ & $1.06\E{14}$ & $9.87\E{13}$\\
      $n_B$ & (fm$^{-3}$) & 0.0767 & 0.0640 & 0.0595 \\
      $\RWS$ & (fm) & 9.94 & 12.75 & 14.86 \\ 
      $R$ & (fm) & 3.90 & 5.63 & 8.05 \\ 
      $n_{n,1}$ & (fm$^{-3}$) & 0.0644 & 0.0541 & 0.0505 \\
      $n_{n,2}$ & (fm$^{-3}$) & 0.0890 & 0.0938 & 0.0950 \\
      $n_{p,2}$ & (fm$^{-3}$) & 0.0075 & 0.0128 & 0.0146 \\
    \end{tabular}
  \end{ruledtabular}
\end{table}
we show the corresponding three sound velocities $v_i = \omega_i/k$ as
functions of the angle $\theta$. The two higher modes are
mixed. Nevertheless one can say that in first approximation the
highest mode corresponds to the longitudinal lattice phonon, i.e., the
displacement $\xiv$ is approximately parallel to $\kv$. Its velocity
depends essentially on $K$ and $\mat{n}_n^b$ and the angle dependence
reflects the anisotropy of $\mat{n}_n^b$.  The second mode is the
superfluid phonon (i.e., the eigenvector is dominated by
$\barphi$). Its angle dependence is surprisingly weak. It turns out
that this is the result of a compensation between the effect of the
angle dependence of the kinetic energy term and the one of the
anisotropic mixing with the longitudinal lattice phonon. Finally, the
lowest-lying phonon is a transverse wave. Without the term $\propto
K_1$ in \Eq{eq:k4-lasagne}, its energy would be approximately
proportional to $k \sin 2\theta$. With the term $\propto K_1$,
however, its energy for $\theta = 90\degree$ remains finite and
proportional to $k^2$, i.e., the sound velocity $v$ depends on
$k$. This is why we show results for two different values of $k$.

\begin{figure}
\includegraphics[width=7cm]{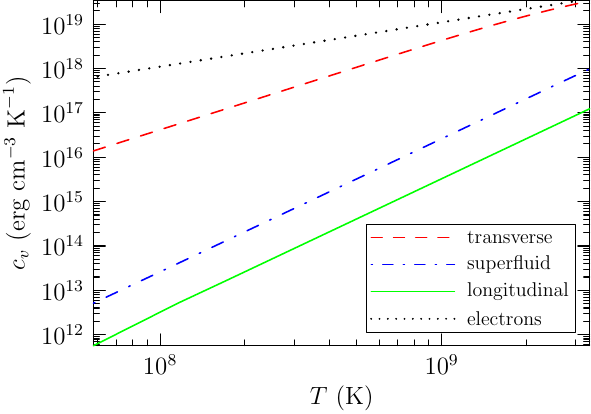}
\caption{\label{fig:cv-lasagne} Temperature dependence of the
  contributions to the specific heat of the lasagne phase
  corresponding to the three phonons shown in
  \Fig{fig:sound-lasagne}. For comparison, the electron contribution 
  $c_{v,e} = \mu_e^2 T/3$ is also shown.}
\end{figure}

We can compute the contribution of each mode $i$ to the specific heat,
\begin{equation}
c_{v, i} = \frac{\partial}{\partial T} \int_{\text{BZ}}
\frac{d^3 k_{\perp}}{(2 \pi)^3} \frac{
  \omega_i(\kv)}{e^{\omega_i(\kv)/T} - 1}\,,
\label{eq:cv-BZ}
\end{equation}
where we have restricted the $\kv$ integral to the first Brillouin
zone (BZ). In the lasagne phase, this means that we integrate $k_x$
and $k_y$ from $-\infty$ to $+\infty$ but restrict the $k_z$ integral
to $|k_z|<\pi/L=\pi/(2\RWS)$. The results are shown in
\Fig{fig:cv-lasagne}.
The specific heat is dominated by the mode which has the lowest
velocity, i.e., the transverse phonon. Furthermore, because of its
complicated angle-dependent dispersion relation, this mode gives rise
to a specific heat proportional to $T^2$ at low temperatures, in
contrast to the usual $T^3$ behavior of the other two
contributions. Notice that a $T^2$ behavior of the specific heat in
the lasagne phase was already found in \Refs{DiGallo2011,Urban2015},
however, the angle dependence of the mode was different there. In any
case, up to $T = 3 \cdot 10^{9}$ K, the electron contribution to
$c_v$, which is linear in $T$, is dominant.

Notice that, if there was not the term $\propto K_1$ that gives a
finite energy to the transverse phonon at $\theta = 90\degree$, its
contribution to the specific heat would diverge. Therefore, $c_v$
depends sensitively on the value of $K_1$. Furthermore, $c_v$ depends
also sensitively on the cut-off $\pi/(2\RWS)$ of the $k_z$
integral. This is a problem since in principle the effective theory
can be assumed to be reliable only at $k\ll 1/\RWS$ and it is
therefore not clear whether one can trust the dispersion relations
$\omega(\kv)$ up to $k=\pi/(2\RWS)$.
\subsection{Spaghetti phase}
\label{sec:resultslasagne}
Although the hexagonal lattice of the spaghetti phase has only a
discrete rotational invariance, the effective Lagrangian is invariant
under rotations around the $z$ axis and we may again assume $k_x =
k\sin\theta$, $k_y = 0$, and $k_z = k\cos\theta$ without loss of
generality. But unlike the lasagne phase, the spaghetti phase supports
shear stress in the $xy$ plane and we have therefore a fourth mode,
namely a transverse mode with $\xiv$ in $y$ direction, which is
decoupled from the other three modes.

\begin{figure}
\includegraphics[width=7cm]{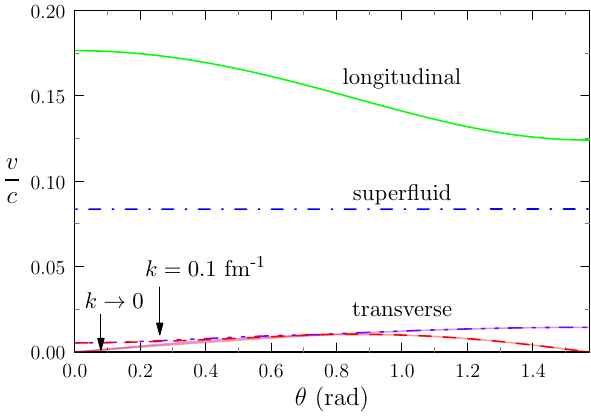}
\caption{\label{fig:sound-spaghetti}Angle dependence of the velocities
  of the four phonons in the spaghetti phase. The parameters are given
  in Table~\ref{tab:parameters}.}
\end{figure}

The corresponding four sound velocities are displayed in
\Fig{fig:sound-spaghetti}.
The two lower modes are transverse lattice phonons, while the two
upper modes are the coupled longitudinal lattice phonon and the
superfluid phonon. Note that, for $\theta = 0$, both transverse modes
are degenerate with $\omega = \sqrt{\smash[b] {K_3/[m (n_{n,xx}^{b} +
      \bar{n}_{p})]}} k^2$, i.e., a $k$ dependent velocity. In the
other limiting case, $\theta = 90\degree$, only the transverse mode
with $\xiv$ in $y$ direction survives (because of the finite shear
modulus $C_{66}$), while the energy of the second transverse mode goes
to zero since a $ z $ independent displacement field in $z$ direction
does not produce any restoring force.

\begin{figure}
\includegraphics[width=7cm]{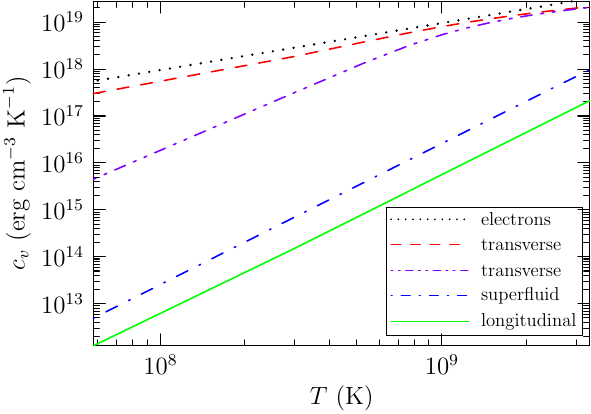}
\caption{\label{fig:cv-spaghetti} Temperature dependence of the
  contributions to the specific heat of the spaghetti phase
  corresponding to the four phonons shown in
  \Fig{fig:sound-spaghetti}.}
\end{figure}

In \Fig{fig:cv-spaghetti}
we display the temperature dependence of the corresponding specific
heat $c_v$. In this case, the first BZ extends from
$-\infty$ to $\infty$ in $z$ direction while in the $xy$ plane it is a
hexagon with side length $b/\sqrt{3}$, which we approximate by a
circle with the same area, i.e., with radius $k_{\perp\max} =
2/\RWS$. Again, the dominant contribution to the heat capacity comes
from the low-lying transverse modes. Because of their different angle
dependence, these contributions behave very differently. The
contribution of the mode with $\xiv$ in the $xz$ plane (or generally,
in the plane spanned by $\kv$ and the $z$ axis) is linear in $T$ and about one half of the electron contribution, while the contribution of the mode with
$\xiv$ in $y$ direction (or perpendicular to the plane spanned by
$\kv$ and the $z$ axis) is much smaller and behaves like $T^{5/2}$ at
low temperature. The contributions of the coupled longitudinal lattice
phonon and superfluid phonon are proportional to $T^3$ and can be
practically neglected in the temperature range of interest.

Note, however, that the warning given at the end of
\Sec{sec:resultslasagne} applies also here.

\subsection{BCC crystal}
\begin{figure}
\includegraphics[width=7cm]{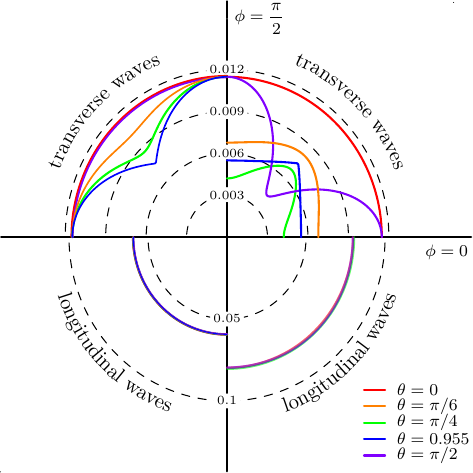}
\caption{\label{fig:sound-crystal} Dependence of the velocities of the
  four phonons in the BCC crystal on the azimuthal angle $\phi$ for
  five different values of the polar angle: $\theta = 0$,
  $30\degree$, $45\degree$, $54.7\degree$, and $90\degree$,
  corresponding to $\cos\theta = 1$, $\sqrt{3}/2$, $1/\sqrt{2}$,
  $1/\sqrt{3}$, and $0$, respectively. The speeds of sound are
  indicated by the radial distance of the curves from the
  origin. Note that the scales are different for the transverse
  (upper scale) and longitudinal (lower scale) modes. The parameters
  are given in Table~\ref{tab:parameters}.}
\end{figure}

In the BCC crystal, we have, as in the spaghetti phase, four
modes. Although they are in principle all coupled with one another
(because of the anisotropy), the two lowest ones are essentially
transverse lattice phonons and only weakly coupled with the two higher
ones corresponding to the mixed longitudinal lattice phonon and
superfluid phonon. Now the speeds of sound depend on the polar angle
$\theta$ and on the azimuthal angle $\phi$. In \Fig{fig:sound-crystal}
we show as an example the $\phi$ dependence of the phonon velocities
for different angles $\theta$. Again, the transverse phonons have a
much lower velocity than the longitudinal ones. The angle dependence
of the transverse modes is very strong, while that of the longitudinal
modes is in practice negligible.

\begin{figure}
\includegraphics[width=7cm]{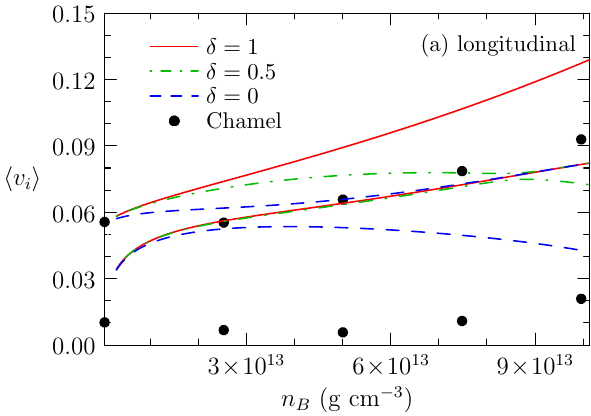}
\includegraphics[width=7cm]{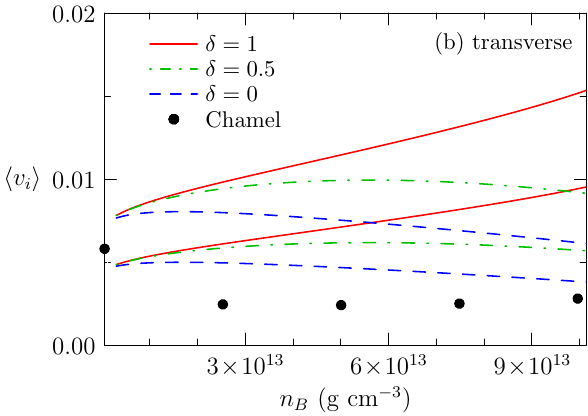}
\caption{\label{fig:density-dependence} Density dependence of the
  angle-averaged velocities of the four phonons in the crystalline
  phase: (a) longitudinal waves, (b) transverse waves. The solid lines ($\delta=1$) 
  correspond to the superfluid hydrodynamic model for the entrainment, \Eq{eq:nnb-crystal},
  while the dashed lines ($\delta = 0.5$, $0$) take into account a possible reduction of
  the superfluid density inside the clusters (cf. \Eq{eq:nnb-crystal-delta}).
  The density dependence of the parameters ($R, \RWS, n_{n,i},
  n_{p,2}$) of the crust were taken from the ETF calculation of
  \Ref{Martin2015}. The points are the results of \Ref{Chamel2013}.}
\end{figure}

In contrast to the lasagne and spaghetti phases, in the crystal all
phonon velocities are already at leading order ($\omega\propto k$)
non-vanishing for all angles. Therefore, it is not necessary to
include higher-order terms (such as $\delta\calV_{\CS,2}$) in the 
Lagrangian. Also, the calculation of the
specific heat is simplified since, at low temperatures where the
effective theory can be assumed to be valid, one may take the integral
in \Eq{eq:cv-BZ} over the whole $\kv$ space instead of the first
BZ. Then one finds
\begin{equation}
  c_{v,i} = \frac{2\pi^2T^3}{15\langle v_i\rangle^3}
\end{equation}
if the mode velocity is angle averaged in a suitable way as
\begin{equation}
  \langle v_i\rangle = \left(\int \!\frac{d\Omega}{4\pi}\,
  \frac{1}{v_i^{3}}\right)^{-1/3}\,,
\end{equation}
where $v_i = \omega_i(\kv)/k$ is the (angle dependent) phonon velocity
and $\Omega$ is the solid angle. The four angle-averaged mode
frequencies are displayed in \Fig{fig:density-dependence}
as the solid lines.

Compared with \Ref{Chamel2013}, except for the highest mode, for which the
agreement is reasonable, the phonon 
velocities are generally much higher in our model. The reason
for this discrepancy is that in the hydrodynamic model, the suppression of
the superfluid neutron density $n_n^s$ due to entrainment is much
weaker and therefore $n_n^b$ is smaller than in the band-structure
calculation \cite{Chamel2012} used in \Ref{Chamel2013}. This results in higher phonon
velocities which are roughly proportional to $(n_n^b+\bar{n}_p)^{-1/2}$ 
(except for the superfluid phonon), see Appendix \ref{app:crystal}. If we
artificially replace our $n_n^s/\bar{n}_n = 1-n_n^b/\bar{n}_n$ from
\Eq{eq:nnb-crystal} by the numerical results of
\Ref{Chamel2012}\footnote{Our $n_n^s/\bar{n}_n$ corresponds to
  $n_n^c/n_n$ in the notation of \Ref{Chamel2012}.}, we obtain a
reasonable agreement with the phonon velocities shown in
\Ref{Chamel2013} in spite of our slightly different crust composition.

The weaker entrainment found in the hydrodynamic model allows
one to solve some problems with the description of glitches
\cite{Martin2016}. It also goes into the same direction as a recent study
based on a completely different approach \cite{Watanabe2017}. Nevertheless, one
might question the validity of the hydrodynamic approach because the
coherence length, especially inside the clusters, is not small enough
compared to the cluster size. To account for this problem in a
pragmatic way, in \Ref{Martin2016} a parameter $0\leq\delta \leq 1$
was introduced which characterizes the fraction of effectively
superfluid neutrons inside the clusters in the sense that the
microscopic neutron current inside the cluster is given by $\delta \,
n_{n,2}\nablav\phi/m+(1-\delta)n_{n,2}\vv_p$ instead of $n_{n,2}
\nablav\phi/m$. Then, the entrainment increases (although it remains
always weaker than that of \Ref{Chamel2012}) and the expression
(\ref{eq:nnb-crystal}) for $n_n^b$ is replaced by
\begin{equation}
  n_n^b = u n_{n,2}
  \Big(1-\delta+\frac{(\delta-\gamma)^2}{(2\gamma+\delta)}\Big)\,.
\label{eq:nnb-crystal-delta}
\end{equation}
In the limiting case $\delta\to 0$ (i.e., all neutrons inside the
clusters flow together with the protons and the neutron gas has to
flow around the clusters), one recovers the result obtained in
\Ref{Sedrakian1996}.

The results discussed so far correspond to the case $\delta=1$. 
In \Fig{fig:density-dependence}, we also display results obtained for
$\delta = 0.5$ and $\delta=0$ (dashed lines). With decreasing $\delta$,
the phonon velocities become smaller, getting somewhat closer to the 
results of \Ref{Chamel2013}. As in \Ref{Chamel2013}, we see an avoided
crossing of the longitudinal lattice phonon and the superfluid phonon 
due to the mixing term. Note that the heat capacity goes like $1/v^3$, so
that the remaining uncertainty of the phonon velocities may change the
specific heat by a huge factor.

\section{Conclusion}
\label{sec:conclusions}

Generalizing the ideas of \Ref{Cirigliano2011} to the pasta phases, we
have constructed the Lagrangian density of an effective theory
describing the long-wavelength dynamics of the inner crust in terms of
the coarse-grained variables $\barphi$ and $\xiv$. Also in the pasta
phases, there is a mixing term between lattice and superfluid phonons,
which is now anisotropic, i.e., angle dependent.

To determine the parameters of the effective theory, we have used the
superfluid hydrodynamics approach of \Ref{Martin2016} for the kinetic
energy including the entrainment, and the phase-coexistence model for
the strong-interaction contribution to the potential energy. The
incompressibility of the lattice comes essentially from the electron
gas. Coulomb and surface energies are responsible for the elastic
properties of the lattice, which we have revisited in detail,
mainly following \Ref{Pethick1998}. But we allow also for compression due to
motion in the direction of the pasta, without change of the distance
between pasta structures [see \Fig{fig:deformation-lasagne}(b)], which
was not considered in \Ref{Pethick1998}.

We have then applied the effective theory to compute the phonon
velocities and specific heats of the lasagne, spaghetti, and BCC
crystal phases. The angle dependence of the phonon velocities is very
strong in all three phases. In the pasta phases, certain transverse phonon
velocities even become equal to zero for $\theta = 0$
or $90\degree$, leading to a qualitative change in the behavior of the
specific heat at low temperatures, as already noticed in
\Refs{DiGallo2011,Urban2015}. In particular, in the spaghetti phase, the
contribution of one of the transverse phonons to the specific heat is
linear in $T$ and about one half of the electron contribution.

The largest uncertainty comes probably from the entrainment
parameters. The superfluid hydrodynamics approach used here predicts a
much weaker entrainment than the band-structure calculation used in
\Ref{Chamel2013}. Therefore, our phonon velocities in the crystalline
phase are higher than those of \Ref{Chamel2013}. Since a full
so-called Quasiparticle Random-Phase Approximation (QRPA) calculation
is only feasible in a WS cell \cite{Khan2005,Inakura2017} but not in
the periodic structure of the inner crust, the superfluid
hydrodynamics approach should be cross-checked with the QRPA in the WS
cell.

Note that in the present work, as in \Refs{DiGallo2011,Martin2016}, we
have neglected the microscopic entrainment related to the dependence
of the microscopic neutron effective mass inside the cluster on the
proton density (and vice versa) \cite{Borumand1996}. However, this
effect seems to be rather weak \cite{Urban2015}.

In the present work, questions related to phonon damping were not
addressed, but they are important in the context of the phonon
contribution to the heat conductivity \cite{AshcroftMermin}. Phonon damping arises from
their coupling to electrons \cite{Kobyakov2017}, whose dynamics is not
yet fully treated, from the phonon scattering off impurities and from
processes involving more than two phonons \cite{Page2012}. To describe
the phonon-phonon coupling, one has to include terms into the
effective Lagrangian involving terms of higher order in the fields
$\barphi$ and $\xiv$.

Finally we note that in the stage of finalizing the present
manuscript, a very similar study by Kobyakov and Pethick appeared
\cite{Kobyakov2018}.

\appendix
\section{Evaluation of the sums in the hexagonal lattice}
\label{app:sumlm}
We consider the hexagonal lattice defined by $\rv_{lm} =
l\vb{a}_1+m\vb{a}_2$, with $\vb{a}_1 = L \ev_x$ and $\vb{a}_2 = (L/2)
\ev_x+(\sqrt{3}L/2) \ev_y$, and $\ev_i$ the unit vector in direction
$i$. The unit cell $A$ is the rhombus defined by the vectors
$\vb{a}_1$ and $\vb{a}_2$ and has an area of $|A|=\sqrt{3}L^2/2 = \pi
\RWS^2$. The reciprocal lattice is given by the vectors $\kv_{lm}$
defined below \Eq{eq:eClasagnelattice}, with $\vb{b}_1 = b\ev_y$ and
$\vb{b}_2 = -(\sqrt{3}b/2)\ev_x+(b/2)\ev_y$. A function $f(\rv)$
($\rv$ denotes here a two-component vector in the $xy$ plane) having
the periodicity of this lattice can be expanded in a Fourier series
$f(\rv) = \sum_{lm} f_{\kv_{lm}} e^{i\kv_{lm}\cdot\rv}$, with
$f_{\kv_{lm}} = (1/|A|) \int_A d^2r f(\rv)e^{-i\kv_{lm}\cdot\rv}$. We
will also make use of the relation
\begin{equation}
  |A| \sum_{lm} f_{\kv_{lm}}g_{-\kv_{lm}}=\int_A d^2r f(\rv)g(\rv)\,.
  \label{eq:sumintegral}
\end{equation}

The lattice is symmetric with respect to rotations by
$60\degree$. Such a rotation changes the indices $m$ and $l$ according
to $(l,m)\mapsto (-m,l+m)$. For a given pair $(l,m)$, let us denote by
$(l^{(n)},m^{(n)})$ the indices one obtains by successively applying
$n$ such rotations, in particular $(l^{(0)},m^{(0)}) =
(l^{(6)},m^{(6)}) = (l,m)$. Suppose one wants to sum a term $a_{lm}$
over $l$ and $m$. Then one can write
\begin{equation}
  \sum_{lm} a_{lm} = \sum_{lm}\frac{1}{6}\sum_{n=0,5} a_{l^{(n)}m^{(n)}}
  \equiv \sum_{lm} a^{(\text{symm})}_{lm}
  \label{eq:symmetrize}
\end{equation}
in order to make the symmetry already apparent before the summation is
performed. This trick has been used to obtain the compact formula for
$C_{66}$ in \Eq{eq:C66lattice}.

Consider now the functions
\begin{multline}
  f_1(\rv) = \theta(R-r)\,,\quad
  f_2(\rv) = r_x \theta(R-r)\,,\\
  f_3(\rv) = (R^2-r^2) \theta(R-r)
\end{multline}
and their Fourier transforms
\begin{multline}
  f_{1,\kv} = \frac{2R}{\RWS^2}\, \frac{J_1(kR)}{k}\,,\quad
  f_{2,\kv} = \frac{2iR^2}{\RWS^2}\,\frac{k_x J_2(kR)}{k^2}\,,\\
  f_{3,\kv} = \frac{4R^2}{\RWS^2}\,\frac{J_2(kR)}{k^2}\,.
\end{multline}
Using \Eq{eq:sumintegral} with $f = g = f_1$, we find
\begin{equation}
\sum_{lm} \frac{[J_1(k_{lm}R)]^2}{k_{lm}^2} = \frac{\RWS^2}{4}\,.
\end{equation}
Similarly, with $f = f_1$, $g = f_3$ we obtain
\begin{equation}
\sum_{lm} \frac{J_1(k_{lm}R)J_2(k_{lm}R)}{k_{lm}^3} = \frac{\RWS^2 R}{16}\,.
\end{equation}
With $f = g = f_2$ and using \Eq{eq:symmetrize} with
$[(k_{lm})_x^2]^{\text{(symm)}} = k_{lm}^2/2$, we obtain
\begin{equation}
  \sum_{lm} \frac{[J_2(k_{lm}R)]^2}{k_{lm}^2} = \frac{\RWS^2}{8}\,.
\end{equation}
These equations allow us to transform \Eq{eq:C66lattice} into
\Eq{eq:C66analytic}.

\section{Matrices}
\label{app:matrices}
Assuming $\barphi = \barphi_0 e^{i (\kv \cdot \rv - \omega t)}$ and
$\xiv = \xiv_0 e^{i (\kv \cdot \rv - \omega t)}$, one can write the
Euler-Lagrange equations (\ref{eq:EulerLagrangephi}) and
(\ref{eq:EulerLagrangexi}) in matrix form as follows:
\newcommand{\alignline}{\vphantom{\dot{\xi}_{y}}}
\begin{equation}
\left(\begin{smallmatrix}
  \alignline \omega & 1 & 0 & 0 & 0 & 0 & 0 & 0 \\
  \alignline A_{21} & \omega & 0 & A_{24} & 0 & A_{26} & 0 & A_{28}  \\
  \alignline 0 & 0 & \omega & 1 & 0 & 0 & 0 & 0 \\
  \alignline 0 & A_{42} & A_{43} & \omega & A_{45} & 0 & A_{47} & 0 \\
  \alignline 0 & 0 & 0 & 0 & \omega & 1 & 0 & 0 \\
  \alignline 0 & A_{62} & A_{63} & 0 & A_{65} & \omega & A_{67} & 0 \\
  \alignline 0 & 0 & 0 & 0 & 0 & 0 & \omega & 1 \\
  \alignline 0 & A_{82} & A_{83} & 0 & A_{85} & 0 & A_{87} & \omega
\end{smallmatrix}\right)
\left(\begin{smallmatrix}
   \alignline -\barphi \\
   \alignline i \dot{\barphi} \\
   \alignline -\xi_{x} \\
   \alignline i \dot{\xi}_{x} \\
   \alignline - \xi_{y} \\
   \alignline i \dot{\xi}_{y} \\
   \alignline - \xi_{z} \\
   \alignline i \dot{\xi}_{z}
\end{smallmatrix}\right)
   = 0.
\end{equation}
\subsection{Lasagne phase}
Assuming without loss of generality that $\kv$ lies in the $xz$ plane,
the matrix elements $A_{ij}$ for the lasagne phase read:
\allowdisplaybreaks
\begin{alignat}{3}
  A_{21} &= \tfrac{\bar{n}_n}{\Gamma m} k_x^2 + \tfrac{n_{n,zz}^s}{\Gamma m}k_z^2\,,
  &&\quad&
  A_{24} &= \tfrac{\gamma_p\bar{n}_p}{\Gamma} k_x\,,
  \nonumber\\
  A_{26} &= 0\,,
  &&\quad&
  A_{28} &= \tfrac{n_{n,zz}^b + \gamma_p \bar{n}_p}{\Gamma} k_z\,,
  \nonumber\\
  A_{42} &= \tfrac{\gamma_p}{m} k_x\,,
  &&\quad&
  A_{43} &= \tfrac{K+B_{11}}{m \bar{n}_p} k_x^2\,,
  \nonumber\\
  A_{45} &= 0\,,
  &&\quad &
  A_{47} &= \tfrac{K+B_{13}}{m \bar{n}_p} k_x k_z\,,
  \nonumber\\
  A_{62} &= 0\,,
  &&\quad&
  A_{63} &= 0\,,
  \nonumber\\
  A_{65} &= 0\,,
  &&\quad&
  A_{67} &= 0\,,
  \nonumber\\
  A_{82} &= \tfrac{n_{n,zz}^b + \gamma_p\bar{n}_p}{m (n_{n,zz}^b+\bar{n}_p)} k_z\,,
  &&\quad&
  A_{83} &= \tfrac{(K + B_{13})}{m (n_{n,zz}^b + \bar{n}_p)} k_x k_z\,,
  \nonumber\\
  A_{85} &= 0\,,
  &&\quad&
  A_{87} &= \tfrac{(K+B_{33})k_z^2+K_1 k_x^4}{m (n_{n,zz}^b + \bar{n}_p)}\,.
\end{alignat}
\subsection{Spaghetti phase}
Again, we may assume without loss of generality that $\kv$ lies in the
$xz$ plane.  Then the matrix elements $A_{ij}$ for the spaghetti phase
read:
\begin{alignat}{3}
  A_{21} &= \tfrac{n_{n,xx}^s}{\Gamma m} k_x^2 + \tfrac{\bar{n}_n}{\Gamma m} k_z^2\,, 
  &&\quad&
  A_{24} &= \tfrac{n_{n,xx}^b + \gamma_{p} \bar{n}_p}{\Gamma} k_x\,,
  \nonumber\\
  A_{26} &= 0\,,
  &&\quad&
  A_{28} &= \tfrac{\gamma_{p} \bar{n}_p}{\Gamma} k_z\,,
  \nonumber\\
  A_{42} &= \tfrac{n_{n,xx}^b + \gamma_{p} \bar{n}_p}{m (n_{n,xx}^b + \bar{n}_p)} k_x\,,
  &&\quad&
  A_{43} &= \tfrac{(K+B_{11}+C_{66}) k_x^2 + K_3 k_z^4}{m (n_{n,xx}^b + \bar{n}_p)}\,,
  \nonumber\\
  A_{45} &= 0\,,
  &&\quad&
  A_{47} &= \tfrac{K+B_{13}}{m (n_{n,xx}^b + \bar{n}_p)} k_x k_z\,,
  \nonumber\\
  A_{62} &= 0\,,
  &&\quad&
  A_{63} &= 0\,,
  \nonumber\\
  A_{65} &= \tfrac{C_{66} k_x^2 + K_3 k_z^4}{m (n_{n,xx}^b + \bar{n}_p)}\,,
  &&\quad&
  A_{67} &= 0\,,
  \nonumber\\
  A_{82} &= \tfrac{\gamma_p}{m} k_z\,,
  &&\quad&
  A_{83} &= \tfrac{K + B_{13}}{m \bar{n}_p} k_x k_z\,,
  \nonumber\\
  A_{85} &= 0\,,
  &&\quad&
  A_{87} &= \tfrac{K+B_{33}}{m \bar{n}_p} k_z^2\,.
\end{alignat}
\subsection{Crystalline phase}
\label{app:crystal}
In the crystal, we have to allow for a general $\kv$ with components $k_x$,
$k_y$, and $k_z$:
\begin{alignat}{3}
  A_{21} &= \tfrac{n_n^s}{\Gamma m} (k_x^2+k_y^2+k_z^2)\,,
  &&\quad&
  A_{24} &= \tfrac{n_n^b + \gamma_p \bar{n}_p}{\Gamma} k_x\,,
  \nonumber\\
  A_{26} &= \tfrac{n_n^b + \gamma_p \bar{n}_p}{\Gamma} k_y\,,
  &&\quad& A_{28} &= \tfrac{n_n^b + \gamma_p \bar{n}_p}{\Gamma} k_z\,,
  \nonumber\\
  A_{42} &= \tfrac{n_n^b + \gamma_p \bar{n}_p}{m (n_n^b + \bar{n}_p)} k_x\,,
  &&\quad&
  A_{43} &= \tfrac{C_{12} k_x^2 + C_{44} (k_y^2+k_z^2)}{m (n_n^b + \bar{n}_p)}\,,
  \nonumber\\
  A_{45} &= \tfrac{C_{11} + C_{44}}{m (n_n^b + \bar{n}_p)} k_x k_y\,,
  &&\quad&
  A_{47} &= \tfrac{C_{11} + C_{44}}{m (n_n^b + \bar{n}_p)} k_x k_z\,,
  \nonumber\\
  A_{62} &= \tfrac{n_n^b + \gamma_{p} \bar{n}_p}{m (n_n^b + \bar{n}_p)} k_y\,,
  &&\quad&
  A_{63} &= \tfrac{C_{11} + C_{44}}{m (n_n^b + \bar{n}_p)} k_x k_y\,,
  \nonumber\\
  A_{65} &= \tfrac{C_{12} k_y^2 + C_{44} (k_x^2+k_z^2)}{m (n_n^b + \bar{n}_p)}\,,
  &&\quad&
  A_{67} &= \tfrac{C_{11} + C_{44}}{m (n_n^b + \bar{n}_p)} k_y k_z\,,
  \nonumber\\
  A_{82} &= \tfrac{n_n^b + \gamma_p \bar{n}_p}{m (n_n^b + \bar{n}_p)} k_z\,,
  &&\quad&
  A_{83} &= \tfrac{C_{11} + C_{44}}{m (n_n^b + \bar{n}_p)} k_x k_z\,,
  \nonumber\\
  A_{85} &= \tfrac{C_{11} + C_{44}}{m (n_n^b + \bar{n}_p)} k_y k_z\,,
  &&\quad&
  A_{87} &= \tfrac{C_{12} k_z^2 + C_{44} (k_x^2+k_y^2)}{m (n_n^b + \bar{n}_p)}\,.
\end{alignat}

\bibliography{mabiblio}

\end{document}